\documentclass[aps,prb,amsmath,amssymb,floatfix,twocolumn]{revtex4}
\pdfoutput=1
\usepackage{graphicx,bm}
\begin{document}

\title{Quantum  oscillations in  graphene in the presence of disorder and interactions}

\author{Pallab Goswami}
\author{Xun Jia}
\author{Sudip Chakravarty}

\affiliation{Department of Physics, University of California Los Angeles,
Los Angeles, CA 90095-1547}
\date{\today}

\begin{abstract}
Quantum oscillations in graphene is discussed.  The effect of interactions are addressed by Kohn's theorem regarding de Haas-van Alphen oscillations, which states that electron-electron interactions cannot affect the oscillation frequencies as long as disorder is neglected and the system is sufficiently screened, which should be valid for chemical potentials not very close to the Dirac point. We determine the positions of Landau levels in the presence of potential disorder from exact transfer matrix and  finite size diagonalization calculations.  The positions are shown to be  unshifted even for  moderate disorder; stronger disorder, can, however, lead to shifts, but this also appears minimal even for disorder width as large as one-half of the bare hopping matrix element on the graphene lattice. Shubnikov-de Haas oscillations of the conductivity  are calculated analytically within a self-consistent Born approximation of impurity scattering.  The oscillatory part of the conductivity follows the widely invoked Lifshitz-Kosevich form when certain mass and frequency parameters are properly interpreted. 
\end{abstract}
 \maketitle

\section{introduction}

Quantum oscillations of magnetization (de Haas-van Alphen effect: dHvA), of conductivity (Shubnikov-de Haas effect: SdH) and that of Hall coefficient ($R_{H}$) are excellent  probes for  topologies of Fermi surfaces and  properties of Fermi liquids.~\cite{Shoenberg:1984}  
Recently  interesting experiments have led to considerable insight into physical systems such as graphene~\cite{Novoselov:2005} and high temperature superconductors.~\cite{Doiron-Leyraud:2007,Chakravarty:2008} These oscillations fundamentally arise from Landau level quantization and hence their frequencies should be robust with respect to electron-electron interaction, crystalline potential, as long as an effective continuum theory exists, and under modest broadening of the Landau levels due to impurity scattering. On the other hand  the waveform (harmonic content)  or amplitudes are sensitive to many details. 

An elegant theorem for a translationally invariant continuum system in two dimensions and short range electron-electron interactions were formulated by Kohn~\cite{Kohn:1961} to reveal the robustness of the frequencies. This is in contrast to the more complex many-particle analysis of Luttinger,~\cite{Luttinger:1961} which, in principle, can also address the waveform or the amplitude of the oscillations, but appears to fail in two dimensions.~\cite{Curnoe:1998}
The reason behind Kohn's theorem is intuitively clear. The magnetic field is never a small perturbation, especially in two dimensions, where the spectrum converts from continuous to discrete in its presence. Thus, the problem must be exactly diagonalized with the magnetic field before considering the perturbative effects of  short-ranged electron-electron interaction, which can in principle be well behaved, modulo a quantum phase transition or a broken symmetry. Moreover, when the Landau levels are completely filled, the ground state is non-degenerate, resulting in a special stability, similar to magic nuclei. As the magnetic field is tuned through a sequence of filled Landau levels, the macroscopic state   repeats itself, hence the periodicity; of course, this is strictly valid only if the relevant Landau level index, $n\gg 1$.

One of the purposes of the present paper is to revisit Kohn's theorem in the context of graphene where SdH oscillation frequencies have been shown to be remarkably close to what a pure noninteracting theory would predict.~\cite{Novoselov:2005} We show that this is a consequence of Kohn's theorem despite the broken translational symmetry due to  the crystalline potential, strengthening further the evidence of the Weyl fermionic character of the excitation spectra.  In the process, we bring out some of the subtle aspects of the theorem and its connection with the more familiar Luttinger theorem regarding the volume of the Fermi surface.~\cite{Luttinger:1960} 

The second purpose is to provide an exact calculation of the Landau levels in the presence of disorder and an analytical  self-consistent calculation of SdH within self-consistent Born approximation (SCBA). A non-self consistent calculation was previously reporrted,~\cite{Sharapov:2004} as well as  a self-consistent calculation for unitary scatterers~\cite{Peres:2006}---perhaps such extreme strong scattering mechanism is not applicable to realistic graphene samples.  Not only do we demonstrate that the oscillation frequencies are unshifted to a good approximation for even moderate disorder, but also obtain the amplitude and the waveform.  The results, when suitably interpreted in terms of a mass parameter and a frequency scale, are similar to the widely used Lifshitz-Kosevich formula,~\cite{Lifshits:1955} which, in a strict sense, cannot be applied in two dimensions.  From our exact numerical calculations, we also show that for very strong disorder the Landau levels {\em are} shifted, but only minimally,  bearing some resemblance to the unitary scatterers treated previously.~\cite{Peres:2006} It is probably true that our results also imply that, to lowest order, results would remain unchanged when both electron-electron interaction and impurity scattering are considered together, at least for weak disorder relevant to experiments.  The extension to longer ranged impurity scattering is straightforward and is not discussed.

The plan of the paper is as follows: in Sec. II we discuss Kohn's theorem and apply it to graphene. In Sec. III we provide exact transfer matrix and diagonalization calculations to determine the position of the Landau levels and their shifts due to impurity scattering. In the second part of this section we calculate SdH oscillations within a self-consistent Born approximation (SCBA) developed by Ando.~\cite{Ando:1974} The brief Sec. IV contains our concluding remarks and there is an Appendix containing calculational details.

\section{Kohn's theorem for quantum oscillations in two dimensions}

\subsection{Nonrelativistic fermions} There is a precise theorem by Kohn that in a two dimensional continuum system electron-electron interaction cannot shift the dHvA  frequency in all orders in perturbation theory.~\cite{Kohn:1961} To clarify its content, it is useful to reconsider it. Let the unperturbed problem be defined by the non-interacting Hamiltonian, $H_{0}$, in a rectangular box $L_{x}\times L_{y}$. In the Landau gauge for the magnetic field $B$ ($c$ being the velocity of light and $m$ the mass of the electrons),
\begin{equation}
H_{0}=\frac{1}{2m}\sum_{i}\left[p_{x,i}^{2}+\left(p_{y,i}+\frac{e}{c}Bx_{i}\right)^{2}\right]
\end{equation} 
The solution to this Landau level problem is of course familiar. The energy eigenvalues and eigenfunctions of an independent electron are 
\begin{equation}
\epsilon_{n,{\bf k}}=\hbar\omega_{c}\left(n+\frac{1}{2}\right), \; \psi_{n, k}= e^{iky}u_{n}\left(x+\frac{\hbar c k}{eB}\right),
\end{equation}
where the frequency $\omega_{c}=\frac{eB}{mc}$ and the degeneracy of each energy level is
\begin{equation}
g=2\frac{\Phi}{\Phi_{0}} ,
\end{equation}
where the total flux threading the system is $\Phi= B L_{x}L_{y}$ and the flux quantum is $\Phi_{0}=hc/e$; the factor of 2 is for spin. 

Even though momenta are not good quantum numbers, it is still useful to depict the spectra on the two-dimensional $k_{x}-k_{y}$-plane. The degenerate spectra, of degeneracy $g$, lie on concentric circles in this plane, separated by $\hbar\omega_{c}$. Since momentum is no longer a good quantum number, the states are not located on specific points on the circle but can be viewed as rotating with frequency $\omega_{c}$. The total number of available states in a volume in the momentum space is unchanged in spite of the Landau quantization.  In particular, if we denote $\Delta A$ to be the area between the concentric circles, then
\begin{equation}
\frac{L_{x}L_{y}}{(2\pi)^{2}}\Delta A=\frac{g}{2}.
\label{eq:degeneracy}
\end{equation}
Imagine that the Fermi level, $\epsilon_{F}$, at $T=0$ is situated on one such concentric level such that all states with energy $E\le  \epsilon_{F}$ are completely  filled and all the levels for $E> \epsilon_{F}$ are completely empty. Then the total  number of occupied states is {\em eaxctly} the same as the system without the magnetic field and the total energy per electron is also {\em exactly} the same. The area enclosed by the Fermi level, $A(\epsilon_{F})$, follows trivially from Eq.~(\ref{eq:degeneracy}) and is
\begin{equation}
2\frac{A(\epsilon_{F})}{(2\pi)^{2}} = \frac{N}{L_{x}L_{y}},
\end{equation}
where $N$ is the total number of particles. That is none other than the Luttinger sum rule. It is important to note that even though a magnetic field is never a small perturbation, the Luttinger sum rule is unchanged.

The magnetic field corresponding to the ground state of a  system with  an integer number, $n(\epsilon_{F})$, of Landau levels completely filled and all the rest completely empty will satisfy
\begin{equation}
\frac{1}{B_{n}}= n(\epsilon_{F})\frac{2\pi e}{\hbar c}\frac{1}{A(\epsilon_{F})}.
\label{eq:Bn}
\end{equation}
Note that  this ground state is an isolated nondegenerate state separated by a gap $\hbar \omega_{c}$ from the excited state. As we increase $B$, the quantized orbits are drawn out of the Fermi level, and sequentially pass through essentially identical set of nondgenerate isolated ground states. This of course results in the periodicity in the properties of the electron gas; the correction in the limit that $n(\epsilon_{F})\gg 1$ is negligible. Periodicity of course does not imply sinusoidal wave form and can contain higher harmonics.

Now, fix $B_{n}$ to a completely filled Landau level and turn on the electron-electron interaction. To all orders in perturbation theory, a nondegenerate isolated ground state will remain nondegenerate and therefore the sequence of states corresponding to fully filled Landau levels as a function of the magnetic field will be the same, as in the noninteracting case. The periodicity is therefore unchanged and is determined by the enclosed area $A(\epsilon_{F})$, which in turn is fixed by the Luttinger sum rule. 

The argument will clearly break down if electron-electron interaction drives a quantum phase transition as a function of the ineteraction strength. A greater likelihood for this happening is when the Landau level is partially filled and degenerate. Indeed, even for higher Landau levels, we know that a zoo of density wave states is a possibility,~\cite{Koulakov:1996} but Kohn's theorem should be robust for fully filled Landau levels. Note that Kohn's theorem makes no statement about the amplitude of oscillations and is also  silent about the waveform of the oscillations. In general the periodicity, when Fourier analyzed will contain harmonics, and the harmonic content of the nonrelativistic case will be different from the relativistic case.

\subsection{Crystalline system: Dirac fermions in graphene}

We now extend Kohn's theorem to graphene. It is known from experiments that SdH oscillation frequencies are in excellent agreement with the noninteracting system, as though electron-electron interaction plays no role. This is as it should be if the theorem holds. There are two basic issues  that need to be dealt with: the crystalline potential that breaks translational invariance and disorder in graphene samples. In this subsection we will discuss the former and will leave the latter to the following section.

As before, first consider the non-interacting problem. The low energy spectrum for which the quasiparticles are well described by continuum 
Lorentz invariant Hamiltonians based on two inequivalent nodes ${\bf K} = (\frac{2\pi}{3a},\frac{4\pi}{3\sqrt{3}a})$ and ${\bf K}'= (\frac{2\pi}{3a},-\frac{4\pi}{3\sqrt{3}a})$ in the tight binding description of the graphene lattice, with $a$ the lattice spacing (cf. below).~\cite{Neto:2008} Near the Dirac point at $\bf K$ the energy eigenfunctions are given by 
\begin{equation}
-i\hbar v_{F}\boldsymbol{\sigma}\cdot\boldsymbol{\nabla}\psi(\mathbf{r}) = E \psi(\mathbf{r})
\end{equation}
where $\boldsymbol \sigma$ are the Pauli matrices and $v_{F}$ is the Fermi velocity.  In the plane wave basis, the $k$-space Hamiltonian $H_{K}=\hbar v_{F}\boldsymbol{\sigma}\cdot {\bf k}$. Similarly, for $K'$, $H_{K'}=\hbar v_{F}\boldsymbol{\sigma}^{*}\cdot {\bf k}$. The spectra for each energy is two-fold degenerate,  ignoring spin. 

The exact energy eigenvalues  for this two-component Weyl fermion  problem are of course trivial, and the corresponding eigenvalues and eigenfunctions in the presence of a perpendicular magnetic field are well known. While the energy eigenvalues differ from the corresponding non-relativistic Landau level problem, the spinor eigenfunctions are easily constructed from the corresponding non-relativistic problem.  However, we will eschew the exact solutions and only take advantage of the validity of the continuum Hamiltonian to formulate the semiclassical dynamics for large Landau levels that leads to the correct quantum  oscillation frequencies. This is exactly what is necessary to demonstrate Kohn's theorem for the general case of electrons in a crystalline environment. We want to show that the Onsager quantization condition follows from the semiclassical quantization of the Landau orbits as the long wavelength  low energy Hamiltonian is a valid description, regardless of whether not the description is in terms of the relativistic Weyl Hamiltonian, as for graphene.  

In general, for  the relativistic  quasiparticles the semiclassical equations of motion of a Bloch electron are different due to the presence of a nontrivial Berry curvature.~\cite{Chang:1996} These are  
\begin{eqnarray}
 \dot{{\bf r}}&=&\frac{1}{\hbar}\boldsymbol{\nabla}_{{\bf k}}E_{\alpha}({\bf k})-\dot{{\bf k}}\times {\bf \Omega}_{\alpha}({\bf k}), \\
\hbar \dot{{\bf k}}&=&-e{\mathbf E}-\frac{e}{c}\dot{{\bf r}}\times {\bf B},
\end{eqnarray}
where ${\bf E}$ and ${\bf B}$ are the electric and magnetic fields respectively. The Berry curvature is defined as
\begin{equation} 
{\bf \Omega}_{\alpha}({\bf k})=\boldsymbol{\nabla}_{{\bf k}}\times {\bf\mathcal{A}}_{\alpha}({\bf k}),
\end{equation}
where 
\begin{equation}
{\bf \mathcal{A}}_{\alpha}({\bf k})=i\langle u_{\alpha}({\bf k})|\boldsymbol{\nabla}_{\bf k}u_{\alpha}({\bf k})\rangle
\end{equation} 
is the Berry vector potential. The periodic part of the Bloch wavefunction is denoted by $u_{\alpha}({\bf k})$, where $\alpha$ is the band index. 

When ${\bf E}=0$ and ${\bf B}=B\hat{z}$, following Chang and Niu,~\cite{Chang:1996} it can be shown that these set of equations of motion lead to the the following semiclassical quantization rule for the area of the orbit in momentum space
\begin{equation}
A_k=\frac{1}{2}\oint({\bf k}\times d{\bf k})\cdot \hat{z}=\frac{2\pi eB}{\hbar c}(n+\gamma).
\end{equation}
and the Maslov index
\begin{equation}
\gamma=\frac{1}{2}-\frac{\Gamma}{2\pi},
\end{equation}
with 
\begin{equation}
\Gamma=\oint {\bf \mathcal{A}}_{\alpha}({\bf k})\cdot d{\bf k} .
\end{equation}
the Berry phase for the orbit. 

This phase easy is to compute for graphene because the eigenspinors for $K$ and $K'$ can be chosen to be
\begin{equation}
u_{\pm,K}=\frac{1}{\sqrt{2}}\left(\begin{array}{c}1 \\\pm e^{i\varphi_{\bf k}}\end{array}\right), \; u_{\pm,K'}=\frac{1}{\sqrt{2}}\left(\begin{array}{c}1 \\\pm e^{-i\varphi_{\bf k}}\end{array}\right)
\end{equation}
where $\varphi_{\bf k}=\tan^{-1}(k_{y}/k_{x})$. A simple computation shows that 
\begin{eqnarray}
{\bf \mathcal{A}}_{K}({\bf k})&=&i\langle u_{\pm, K}({\bf k})|\boldsymbol{\nabla}_{\bf k}u_{\pm,K}({\bf k})\rangle=-\frac{\hat{\theta}}{2 k}, \\
{\bf \mathcal{A}}_{K'}({\bf k})&=&i\langle u_{\pm, K'}({\bf k})|\boldsymbol{\nabla}_{\bf k}u_{\pm,K'}({\bf k})\rangle=\frac{\hat{\theta}}{2 k}.
\end{eqnarray}
where $\hat{\theta}$ is the azimuthal angle in the $k_{x}-k_{y}$ plane and $k=\sqrt{k_{x}^{2}+k_{y}^{2}}$. The Berry phase  $\Gamma$ is therefore $\pm \pi$. 

The semiclassical quantization formula is then
\begin{equation}
A_k=\frac{2\pi eB}{\hbar c} n .
\end{equation}
This implies once again that the magnetic field corresponding to a fully filled Landau level at the Fermi energy is
given {\em exactly} by Eq.~(\ref{eq:Bn}). 
The only remaining subtlety now is to note that the area of the Fermi pocket 
is
\begin{equation}
\pi k_{F}^{2}=\frac{2\pi e B}{\hbar c} n(\epsilon_{F}),
\end{equation}
from which it immediately follows that the Fermi energy for the relativistic Dirac spectrum 
 is given by
\begin{equation}
\epsilon_{F}=\hbar v_{F}k_{F} = \frac{v_{F}}{\ell_{B}}\sqrt{2n(\epsilon_{F})}.
\end{equation}
The magnetic length 
\begin{equation}
\ell_B=\sqrt{\hbar c/(eB)}.
\end{equation}
The appearance of this Berry phase $\Gamma=\pm \pi$ for massless nodal quasiparticles with linear spectrum is  a well known topological property of the Hamiltonian.~\cite{Volovik:2003,Manes:2007}

From here on the argument proceeds identically to the non-relativistic case, the only difference being the spacing between the Landau level, which is not constant as a function of $n$. As we turn on the electron-electron interaction, the energy levels are drawn out of the Fermi level in exactly the same sequence as the non-interacting case. Therefore the frequency of dHvA is unchanged. To the extent that the low energy continuum theory adequately describes graphene, there should be no effect of electron-electron interaction on the  oscillation frequency.

Due to the relativistic spectrum of massless quasiparticles in $(2+1)$-dimensions the density of states vanish linearly  at the nodal points, and the Coulomb interaction is not screened. This leads to strong long range interaction between  the quasiparticles in graphene. In the presence of a moderate nonzero chemical potential the interaction will be screened, however. For small chemical potentials long range interactions can lead to broken symmetries and hence to a failure of Kohn's theorem.   

The nodal relativistic spectra can also arise from a condensate. An example is nodal fermionic   quasiparticles of a particle-hole condensate in $l =2$ angular momentum channel, as in a singlet $d$-density wave (DDW), staggered flux phase, or an orbital antiferromagnet.~\cite{Chakravarty:2001} This linearized Dirac fermion theory is valid for momentum small compared to inverse lattice spacing and energy small compared to the bandwidth. For a wide range of the chemical potential,  small compared to the bandwidth, one can still use the linearized continuum theory and results identical to those above hold.

\section{Effects of disorder}

\subsection{Landau levels in graphene}
\begin{figure}[htb]
    \centering
  \includegraphics[width=7.5cm]{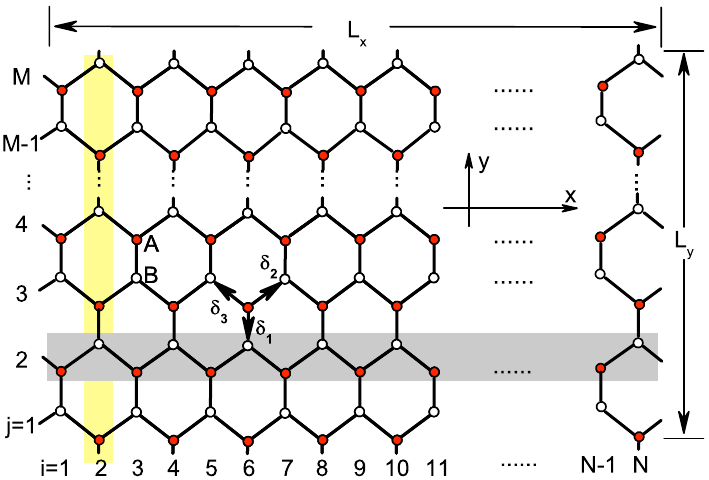}\\
  \caption{(Color online) Graphene lattice. Lattice sites belonging to the
  same shaded horizontal or vertical stripes share the same indices $i$ or $j$.
  The solid circles correspond to the $A$ sublattice and the open circles to the
  $B$ sublattice. The three nearest neighbor vectors joining the two sublattices
  are $\boldsymbol{\delta}_{k}$, $k=1, 2, 3$.}
  \label{lattice}
\end{figure}

Consider the tight-binding Hamiltonian, $H$, on a honeycomb lattice of
dimension $L_x\times L_y$ in a perpendicular magnetic field, as
shown in Fig.~\ref{lattice}, which  is
\begin{equation}\label{eq:ham}
\begin{split}
    H= & \sum_{{\bf n}} (\epsilon_{{\bf n},A} c_{{\bf n},A}^\dagger c_{{\bf n},A} +
    \epsilon_{{\bf n},B} c_{{\bf n},B}^\dagger c_{{\bf n},B})\\
    &- \sum_{{\bf n}}\sum_{k=1}^{3}
    (t_{{\bf n},k}\mathrm{e}^{\mathrm{i} a_{{\bf n},k}}c_{{\bf n},A}^\dagger
    c_{{\bf n}+\boldsymbol{\delta}_k,B}+\mathrm{h.c.}),
\end{split}
\end{equation}
where the summation ${\bf n}$ ranges over all unit cells, and
$c_{{\bf n},A}$, $c_{{\bf n},B}$ are the fermionic annihilation
operators in the unit cell ${\bf n}$ for the sublattices $A$ and
$B$, respectively.  The spin degrees of freedom are omitted, as we
assume that the magnetic field is sufficiently strong to completely
polarize them. In principle, disorder can take various forms,~\cite{Jia:2008} but here we shall consider the on-site energy $\epsilon_{{\bf n},A}$ and
$\epsilon_{{\bf n},B}$ to be random variables uniformly distributed in the range $[-g_V/2, g_V/2]$ 
corresponding to potential disorder. We choose the hopping matrix element
$t_{{\bf n},k}=t=1.0$, 
providing a natural energy scale. The phases
$a_{\vec{n},k}$  are chosen such that the magnetic flux per
hexagonal plaquette, $\phi$, is $1/Q$, in units of the flux quanta
$\phi_0=hc/e$. We choose a gauge such that $a_{{\bf n},1}=\pi i/Q$
for the vertical bonds in slice $i$ as in Fig.~\ref{lattice}, and
$a_{{\bf n},2}=a_{{\bf n},3}=0$.

The Hall conductance $\sigma_\mathrm{xy}$ can be computed by  imposing
periodic boundary conditions in both directions of the system and diagonalizing the 
Hamiltonian~ (\ref{eq:ham})  to obtain a set of
energy eigenvalues $E_\alpha$ and eigenvectors $|\alpha\rangle$ for $\alpha=1,\ldots,L_x\times L_y$.
Then from the Kubo formula:
\cite{Thouless:1982}
\begin{equation}\label{kubo}
    \sigma_\mathrm{xy}(E)=\frac{4\pi ie^2}{L_xL_yh}\sum_{E_\alpha < E < E_\beta}
    \frac{\langle \alpha|v_y|\beta\rangle\langle\beta|v_x|\alpha\rangle-
  ( x\leftrightarrow y)}{(E_\alpha-E_\beta)^2},
\end{equation}
where $v_x=[H,x]/i\hbar$ is the velocity operator along the $x$
direction and  similarly for $v_y$. Note that the bonds in
Fig.~\ref{lattice} that are not parallel to $y$ direction contribute
to both $v_x$ and $v_y$. The summation corresponds to all states
below and above the energy $E$. Finally, the
expression is disorder averaged.

The longitudinal conductance $\sigma_\mathrm{xx}$ is studied using
 the well developed transfer matrix method. Consider a
quasi-1D system, $L_x\gg L_y\equiv 2M$ with a periodic boundary
condition only along the $y$ direction, where $M$ denotes the number
of unit cells in a slice along $y$ direction. Let $\Psi_i =
(\psi_{i,1},\psi_{i,2}, \ldots, \psi_{i,2M})^T$ be the amplitudes
on the slice $i$ for an eigenstate with a given energy $E$; then the amplitudes on the successive slices are related
by the matrix multiplication:
\begin{equation}
    \label{transfermatrix}
    \left[
      \begin{array}{c}
        \Psi_{i+1} \\
        \Psi_{i} \\
      \end{array}
    \right] = \left[
                \begin{array}{cc}
                  V^{-1}_i(E-H_i) & -V_i^{-1}V_{i-1} \\
                  1 & 0 \\
                \end{array}
              \right]\left[
                       \begin{array}{c}
                         \Psi_i \\
                         \Psi_{i-1} \\
                       \end{array}
                     \right],
\end{equation}
where $V_i$ is a diagonal matrix with elements
$(t_{i,1},t_{i,2},\ldots,t_{i,2M})$ representing the hopping matrix
elements connecting the slices $i$ and $i+1$, and $H_i$ is the
Hamiltonian within the slice. All postive Lyapunov exponents of the
transfer matrix,~\cite{Kramer:1996}
$\gamma_1>\gamma_2>\ldots>\gamma_{2M}$, are computed by iterating
Eq.~(\ref{transfermatrix}) and performing orthonormalization
regularly. The convergence of this algorithm is guaranteed by the
well known Osledec theorem.~\cite{Oseledec:1968} The
conductance per square, $\sigma_\mathrm{xx}$, is given by the Landauer
formula\cite{Fisher:1981,Baranger:1989,Kramer:1993,Sheng:2000}(note the factor of $\sqrt{3}$ in the argument of $\cosh$ due to the honeycomb lattice):
\begin{equation}\label{landauer}
    \sigma_\mathrm{xx}=\frac{e^2}{h}\sum_{i=1}^{2M} \frac{1}{\cosh^2
    (2\sqrt{3}M\gamma_i)}.
\end{equation}
The localization length in the
quasi-1D system with  $L_y=2M$ is given by
$\lambda_M = 1/\gamma_{2M}$. 

\begin{figure}[htb]
    \centering
  \includegraphics[width=8.5cm]{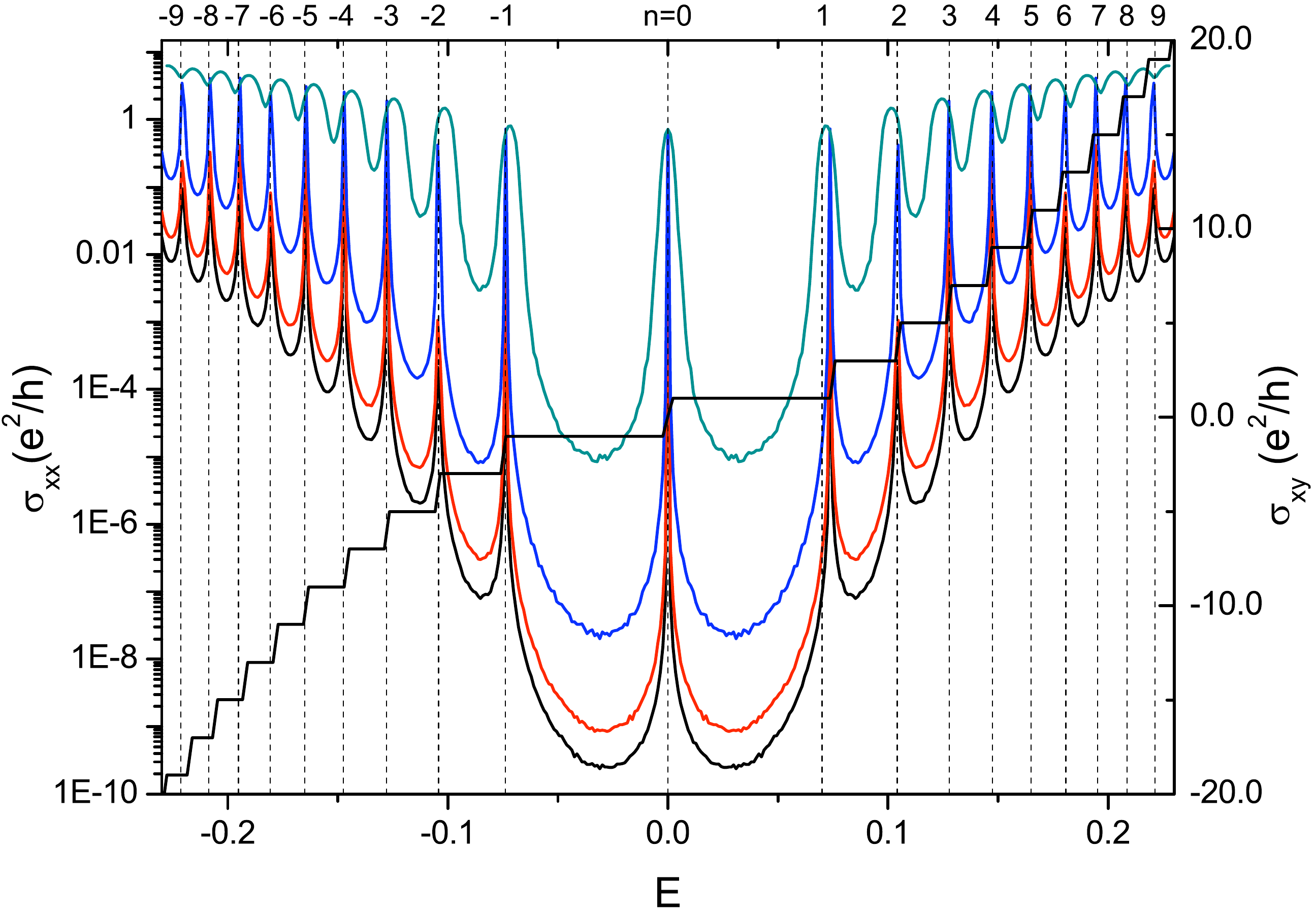}\\
  \caption{(Color online) The Hall conductance $\sigma_\mathrm{xy}$ (plateaus) and
  the longitudinal conductance $\sigma_\mathrm{xx}$ as a function of
  energy $E$. The magnetic flux through the hexagonal plaquette, $\phi=1/2000$
  in units of the flux quantum $\phi_0=hc/e$ and the potential disorder $g_V=0.5, 0.05,0.01,0.001$ from top to bottom.
  For $\sigma_{xx}$, M was chosen to be 48, and, for $\sigma_{xy}$, the system size was chosen to be $4000\times 10$ for the exact diagonalization. In physical units the magnetic field corresponds to $B\approx \mathrm{40 T}$.
  The vertical dashed lines  indicate the locations of
  the Landau levels at $E_n=\textrm{sgn}(n)(\sqrt{3}\pi|n|/1000)^{1/2}$ for the parameters considered here. There are virtually no shifts of the positions of the Landau levels except for $g_{V}=0.5$ for which it is minimal.}
  \label{sigmaxxxy}
\end{figure}

\begin{figure}[htb]
    \centering
  \includegraphics[width=8.5cm]{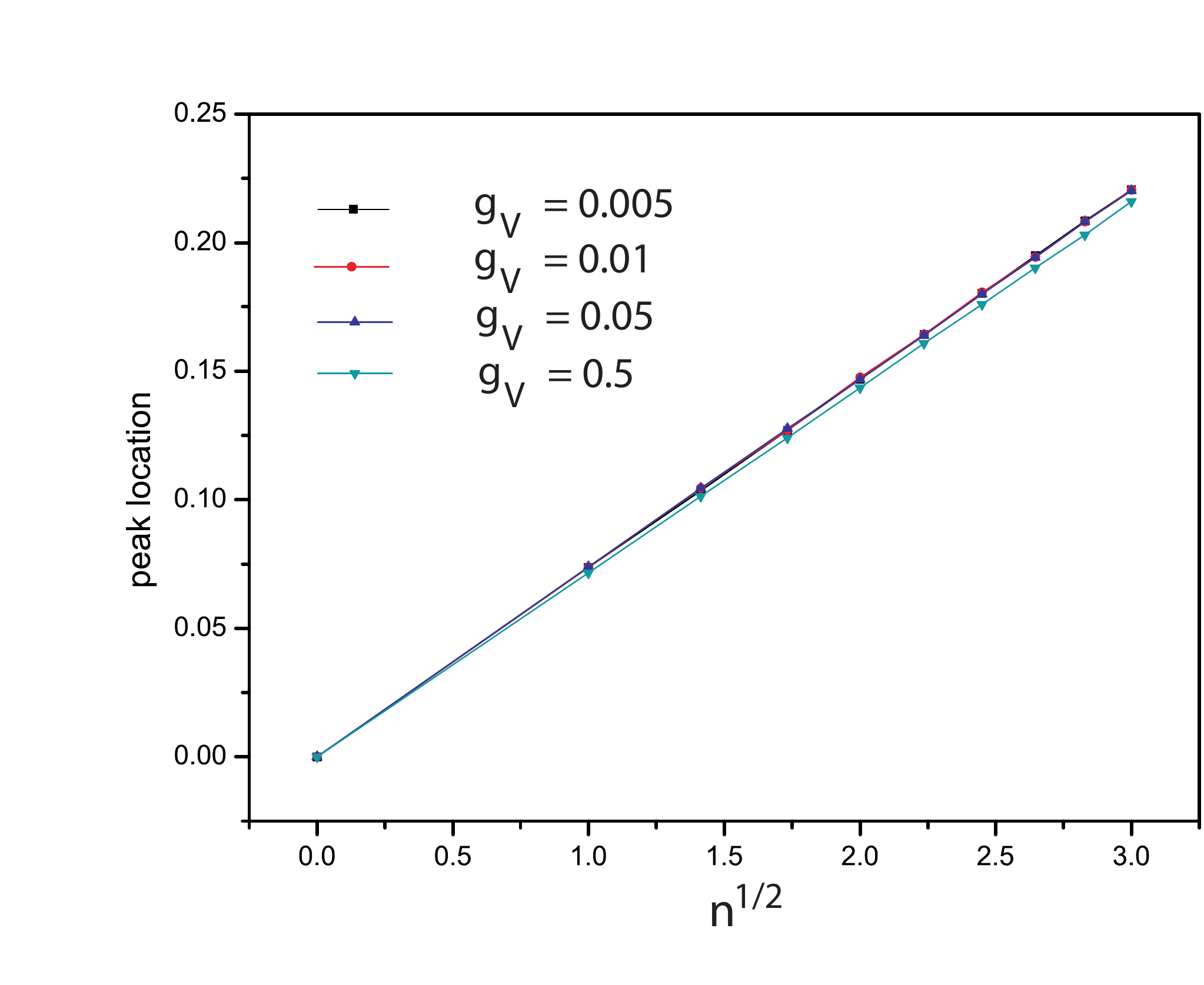}\\
  \caption{(Color online) The location of the peaks in Fig.~\ref{sigmaxxxy}. Note that the slope is essentially the same 
  as the noninteracting system except for $g_{V}=0.5$ for which it is slightly different. }
  \label{peak}
\end{figure}

The results are shown  in
Fig.~\ref{sigmaxxxy}, where the Hall conductance
$\sigma_\mathrm{xy}$ and the longitudinal conductance
$\sigma_\mathrm{xx}$ are computed as a function of energy $E$ with
parameters $\phi=1/2000$ and for  potential disorder
$g_V=0.5, 0.05, 0.01, 0.001$. The longitudinal conductance $\sigma_\mathrm{xx}$ peaks
almost exactly at the Landau levels 
\begin{equation}
E_{n}= {\textrm{sgn}}(n) \hbar  \left(\sqrt{2} \frac{v_{\textrm F}}{\ell_{B}}\right)\sqrt{|n|}, 
\end{equation}
where 
\begin{equation}
v_{\textrm F}=3ta/2\hbar
\end{equation}
is the fermi velocity, $a$
being the lattice spacing. The only exception is the case $g_{V}=0.5$, as shown in Fig.~\ref{peak}, for which there is a minimal shift. Recall that this is very large disorder as the energy scale is in terms of the hopping parameter, $t$, which is set to unity. Each level is broadened due to disorder
and finite size effects. Because the spacings between successive
Landau bands become smaller as $|n|$ increases, the overlap between
the neighboring bands increases,
resulting in the overall parabolic shape of the background
$\sigma_\mathrm{xx}$. The Hall conducivity $\sigma_\mathrm{xy}$
jumps by $2e^2/h$ every time the energy passes through a Landau
level, corresponding to the two fold valley degeneracy in this
system. 

\subsection{SdH Oscillations: analytical results in the self-consistent Born approximation}

Shubnikov-de Haas oscillations in graphene can be calculated numerically by the method described in the previous subsection. However, for realistic fields, it is difficult to control  the accuracy because of the essential singularity corresponding to the magnetic field corresponding to all quantum oscillation phenomena. Moreover, the dependence on the physical parameters are not particularly transparent. For this reason we adopt an analytical SCBA developed by Ando.~\cite{Ando:1974}

In the presence of disorder, the imaginary part of the self energy at the Fermi level is non-zero. Hence, Luttinger's theorem and Kohn's argument are not immediately applicable. However, for weak disorder, the quasiparticle lifetime at the fermi energy, $\tau(\epsilon_{F})\equiv \tau$ (henceforth by $\tau$ we will mean $\tau(\epsilon_{F}$)), can be long and the Fermi surface can be  reasonably well defined within the uncertainty  $\hbar/\tau$. If  the Fermi energy  $\epsilon_F$ is very large, we can use $\hbar/\epsilon_F \tau$ as a small parameter. Due to disorder the Landau levels are broadened into bands and in the limit of overlapping Landau levels ($\omega_c \tau \lesssim1$) explicit calculations using SCBA show that the period of oscillations is unchanged to  an  excellent approximation. The correction terms to the period is of the order $[\hbar/(\epsilon_F \tau)]^{2}$. For unitary scatterers, the shift is much larger,~\cite{Peres:2006} but it is doubtful that such strong potential scattering is relevant to graphene.~\cite{Novoselov:2005}

 Since the gap between the relativistic Landau levels decreases for higher Landau levels, disorder will have stronger effect  in contrast to the equally spaced non-relativistic Landau levels. Hence, in the semiclassical limit of higher Landau levels, the damping  will be stronger for the relativistic case. For the sake of simplicity consider the  short range impurity potential. Following Ando,~\cite{Ando:1974,Zheng:2002} the self-consistency equation for the self-energy $\Sigma(\epsilon)$ in SCBA can be written as
\begin{equation}
\Sigma(\epsilon)= \frac{\hbar^3 \omega^2}{4\pi \epsilon_F \tau}\sum_{m=-N_{c}}^{N_c}\frac{1}{\epsilon-E_{n} -\Sigma(\epsilon)}.
\label{scba1}
\end{equation}
Note that we have introduced a cutoff $N_{c}$, which is of the order of the bandwidth. It determines the  limit of   the applicability of the linearized Dirac spectra.
Let  
\begin{eqnarray}
\epsilon-\Sigma(\epsilon)&=&X(\epsilon)=X'+iX''(\epsilon), \\
a(\epsilon)&=&X^{'}(\epsilon)/(\hbar \omega),\\
b(\epsilon)&=&X^{''}(\epsilon)/(\hbar \omega),
\end{eqnarray}
where $\omega=\sqrt{2}v_{F}/l_{B}$ .
Ignoring the arguments of $a$ and $b$, the self-consistency equation becomes
\begin{equation}
\frac{\epsilon}{\hbar \omega_{c}}-a-ib=\frac{\hbar}{2\pi\epsilon_{F}\tau}\left[\sum_{m=0}^{N_c} \frac{(a+ib)}{(a+ib)^2-m}-\frac{1}{2}\frac{1}{a+ib}\right]
\label{scba2}
\end{equation}

The sum on the right hand side can be calculated using the Poison summation formula described in Appendix~\ref{app:B}, and, to leading orders in $(\hbar/\epsilon_{F}\tau)$,   $a$ and $b$ are given by
\begin{eqnarray}
a&\approx& \frac{\epsilon}{\hbar \omega}\bigg[1-\left(\frac{\hbar}{2\epsilon_F \tau}\right)^2\nonumber\\ & &- \frac{\hbar}{\epsilon_F \tau} e^{-\frac{\pi}{\omega^{*}_{c}\tau}(\frac{\epsilon}{\epsilon_{F}})^{2}}\sin\left(\frac{2\pi y F}{B}+\phi \right)\bigg] \label{scba3}\\
b&\approx&  \frac{\epsilon}{2\epsilon_F \omega \tau}\bigg[1+2e^{-\frac{\pi}{\omega^{*}_{c}\tau}(\frac{\epsilon}{\epsilon_{F}})^{2}}
\cos\left(\frac{2\pi y F}{B}+\phi \right)\bigg] \label{scba4}
\end{eqnarray}
where $\phi \approx \tan^{-1}(\hbar/2\epsilon_F \tau)$, and $y=1-3(\hbar/2\epsilon_F \tau)^2$. 
We have defined a mass parameter $m^{*}=\epsilon_{F}/v_{F}^{2}$, and  $\omega_{c}^{*}=eB/m^{*}c$, the cyclotron frequency of a hypothetical nonrelativistic system. Here the frequency $F$ is 
\begin{equation}
F(\epsilon)= \frac{\hbar c}{2\pi e} A(\epsilon),
\end{equation}
and the area of the Fermi pocket is given by
\begin{equation}
A(\epsilon) = \pi \left(\frac{\epsilon}{\hbar v_{F}}\right)^{2}.
\end{equation}
The condition $a\gg b$ implies $(\hbar/2\epsilon_F \tau)\ll1$, which justifies the weak disorder approximation. When $(\hbar/2\epsilon_F \tau)\sim1$, the SCBA is insufficient to treat the randomness correctly. Hence, in our calculation we assume $(\hbar/2\epsilon_F \tau)\ll1$.

Using the Kubo formula the longitudinal dc conductivity $\sigma_{xx}$ becomes
\begin{eqnarray}
\sigma_{xx}=-\int_{-\infty}^{\infty}d\epsilon \frac{\partial f(\epsilon)}{\partial \epsilon} K_{xx}(\epsilon)
\label{scba5}
\end{eqnarray}
where $f(\epsilon)$ is the Fermi-Dirac distribution function and the kernel $K_{xx}$ is~\cite{Ando:1974}
\begin{eqnarray}
K_{xx}(\epsilon)=\frac{(e\hbar \omega)^2}{\pi^2 \hbar}\sum_{m=0}^{\infty}{\Im} g_m(\epsilon+i0){\Im} g_{m+1}(\epsilon+i0)
\label{scba6}
\end{eqnarray}
In the above equation
\begin{equation}
g_m(\epsilon)=\frac{\epsilon-\Sigma}{(\epsilon-\Sigma)^2-m(\hbar \omega)^{2}}
\end{equation}
At $T=0$, $\sigma_{xx}(T=0)=K_{xx}(\epsilon_F)$. After using the Poisson's summation formula the conductivity can be written as
\begin{widetext}
\begin{eqnarray}
\sigma_{xx}\approx \tilde{\sigma}\left[1+2e^{-\pi/\omega_{c}^{*}\tau} \left \{ \cos\left(\frac{2\pi F(\epsilon_{F})}{B}\right)
-\frac{(\omega_{c}^{*}\tau)^2-1}{(\omega_{c}^{*}\tau)^2+1} \cos\left(y\frac{2\pi F(\epsilon_{F})}{B}+\phi\right)-\frac{3\hbar}{2\epsilon_F \tau}\sin\left(y\frac{2\pi F(\epsilon_{F})}{B}+\phi\right)\right \}\right]
\label{scba7}
\end{eqnarray}
\end{widetext}
where 
\begin{equation}
\tilde{\sigma}=\frac{\sigma_0}{(\omega_{c}^{*}\tau)^2+1},
\end{equation}
and $\sigma_0=4n_ee^2\tau/\pi m^*$; $n_{e}$ is the density of quasiparticles corresponding to one of the valleys for a given spin direction.  

In contrast to the non-relativistic Landau levels,  the frequency of the oscillation is changed by a factor $y$ and there is a small phase shift in the oscillations.  Hence, the robustness of the oscillation frequency and the phase of the oscillations do depend on on disorder but only very weakly. In the SdH experiments, this phase shift has not been observed and this can be attributed to the weakness of the disorder.
When the disorder is weak i.e, $(\hbar/2\epsilon_F \tau)\ll1$, $y\approx1$,  $\phi \approx 0$, and the conductivity becomes
\begin{equation}
\sigma_{xx}\approx \tilde{\sigma}\left[1+\frac{4(\omega_{c}^{*}\tau)^2e^{-\pi/\omega_{c}^{*}\tau}} {(\omega_{c}^{*}\tau)^2+1} \cos\left(\frac{2\pi F(\epsilon_{F})}{B}\right)\right]
\label{scba8}
\end{equation}
If $\omega^{*}_{c}\tau<1$ the oscillations will be heavily damped. 

Although this expression appears identical to the corresponding nonrelativistic formula,~\cite{Ando:1974,Lifshits:1955} it is actually different. The definition of $\omega_{c}^{*}$ depends on the $m^{*}$ that we have defined, which in turn depends on $\epsilon_{F}$. Therefore,  the oscillations will be damped more strongly for higher Landau levels. The reason behind this is the smaller gaps between the higher Landau levels, as compared to the nonrelativistic case.

\section{Conclusions}

In conclusion, we reiterate that the positions of the Landau levels, as well the frequencies of SDH oscillations, are remarkably robust with respect both impurity scattering and electron-electron interaction (as long as the chemical potential is not too close to the Dirac points). This seems to be consistent with experiments.~\cite{Novoselov:2005}
In the presence of both interaction and randomness the Luttinger's many body formalism~\cite{Luttinger:1961} becomes applicable, even  for two dimensions in the presence of disorder and/or thermal broadening. From the above considerations, if the disorder is small enough ($\hbar/\epsilon_{F}\tau\ll 1$), the oscillation frequencies can be expected to be proportional to the true Fermi surface area of the pure interacting problem. In principle one can attempt to treat the interaction within a Hartree-Fock (HF) approximation to estimate the effects of the interaction on the amplitude of the oscillations. But, there is no reason to trust the HF results. The correlation energy contributions can have equally important effects on the amplitude, in particular, on the effective mass parameter. One also needs to account for the disorder induced vertex corrections for the self-energy contributions from the interaction. Such a calculation will be tedious but important to understand the effects of inelastic scattering rates and we shall relegate such detailed calculations for a future publication. 

\begin{acknowledgments}
This work  was supported by NSF under the Grant DMR-0705092. We thank Antonio Castro Neto for helpful discussions. We thank Maria A. H. Vozmediano for pointing out to us Ref.~\onlinecite{Manes:2007}.
\end{acknowledgments}

\appendix

 \section{\label{app:B}SCBA}
Consider the real and the imaginary parts of the Eq. ~(\ref{scba2}) :
\begin{eqnarray}
\frac{4\pi \epsilon_F \tau}{\hbar}\left(\frac{\epsilon}{\hbar \omega}-a\right)+\frac{a}{a^2+b^2}&=&2a\mathcal{S}_{-}\label{scbaA1},\\
\frac{4\pi \epsilon_F \tau}{\hbar}+\frac{1}{a^2+b^2}&=& 2\mathcal{S}_{+},
\label{scbaA2}
\end{eqnarray}
where
\begin{eqnarray}
\mathcal{S}_{\pm}=\sum_{m=0}^{N_{c}}\frac{a^2+b^2\pm m}{(a^2-b^2-m)^2+4a^2b^2} \\
 =\int_{0}^{N_{c}}dm [1+2\sum_{k=1}^{\infty}\cos(2\pi k m)] \nonumber \\
 \times \frac{a^2+b^2\pm m}{(a^2-b^2-m)^2+4a^2b^2}
\end{eqnarray}

Now choosing $m-a^2+b^2=z$, we obtain
\begin{eqnarray}
\mathcal{S}_{+}&=&\int_{-a^{2}+b^{2}}^{N_{c}-a^{2}+b^{2}}dz [1+2\sum_{k=1}^{\infty}\cos(2\pi k(a^2-b^2+z))]\nonumber \\
&&\times \frac{2a^2+z}{z^2+4a^2b^2},\\
\mathcal{S}_{-}&=&\int_{-a^{2}+b^{2}}^{N_{c}-a^{2}+b^{2}}dz [1+2\sum_{k=1}^{\infty}\cos(2\pi k(a^2-b^2+z))]\nonumber \\
&&\times \frac{2b^2-z}{z^2+4a^2b^2}.\\
\end{eqnarray}

Because $N_{c}\gg a(\epsilon)^{2}\gg b(\epsilon)^{2}$ for $\epsilon \lesssim \epsilon_{F}$, we can let the upper limit tend to $\infty$ and the lower limit to $-\infty$. Note that after this change of limits the terms that are odd in $z$ vanish.  We find the following analytic expressions:
\begin{eqnarray}
\mathcal{S}_{+}&=&\frac{\pi a}{b}+\frac{2\pi}{b}\sqrt{a^2+b^2}\sum_{k=1}^{\infty}e^{-4\pi kab}\nonumber \\
&&\times \cos\left[2\pi k(a^2-b^2)+\tan^{-1}\left(\frac{b}{a}\right)\right],
\end{eqnarray}
\begin{eqnarray}
\mathcal{S}_{-}&=&\frac{\pi b}{a}+\frac{2\pi}{a}\sqrt{a^2+b^2}\sum_{k=1}^{\infty}e^{-4\pi kab}\nonumber \\
&&\times \sin\left[2\pi k(a^2-b^2)+\tan^{-1}\left(\frac{b}{a}\right)\right].
\end{eqnarray}
Substituting $\mathcal{S}_{\pm}$ in the Eqs.~(\ref{scbaA1})and (\ref{scbaA2}), retaining only the first harmonic and only terms to leading order in $(\hbar/\epsilon_F \tau)^{2}$ we arrive at Eqs.~(\ref{scba3}) and (\ref{scba4}).

The conductivity kernel $K_{xx}$ in Eq.~(\ref{scba6}) can be expressed as
\begin{eqnarray}
K_{xx}= \frac{e^2b^2}{\pi^2\hbar}\sum_{m=0}^{N_{c}}\frac{a^2+b^2+m}{(a^2-b^2-m)^2+4a^2b^2}\nonumber\\
\times \frac{a^2+b^2+m+1}{(a^2-b^2-m-1)^2+4a^2b^2}
\label{scbaA3}
\end{eqnarray}
  Now using Poisson's summation formula we get
\begin{eqnarray}
K_{xx}&=&\frac{2e^2b^2}{\pi^2 \hbar}[I_1+\sum_{k=1}^{\infty}\cos(2\pi k(a^2-b^2))I_2(k)\nonumber \\
& &-\sum_{k=1}^{\infty}\sin(2\pi k(a^2-b^2))I_3(k)]
\label{scbaA4}
\end{eqnarray}
where
\begin{equation}
2I_1=\int_{-\infty}^{\infty}dz \frac{(2a^2+z)(2a^2+z+1)}{[z^2+4a^2b^2][(z+1)^2+4a^2b^2]}
\end{equation}
\begin{equation}
I_2(k)=\int_{-\infty}^{\infty}dz \frac{\cos(2\pi kz)(2a^2+z)(2a^2+z+1)}{[z^2+4a^2b^2][(z+1)^2+4a^2b^2]}
\end{equation}
\begin{equation}
I_3(k)=\int_{-\infty}^{\infty}dz \frac{\sin(2\pi kz)(2a^2+z)(2a^2+z+1)}{[z^2+4a^2b^2][(z+1)^2+4a^2b^2]}
\end{equation}
We can clearly see $2I_1=I_2(k=0)$, and all the integrals can be evaluated by performing a single integral
\begin{equation}
I_4=\int_{-\infty}^{\infty}dz \frac{e^{i2\pi kz}(2a^2+z)(2a^2+z+1)}{[z^2+4a^2b^2][(z+1)^2+4a^2b^2]}
\end{equation}
The integrand has simple poles at $z=\pm2iab$ and $z=-1\pm 2iab$, and we evaluate $I_4$ by closing the contour in the upper half-plane. It turns out that $I_3=Im I_4=0$ and
\begin{equation}
I_2(k)=Re I_4=\frac{8\pi a}{b}e^{-4\pi kab}\frac{a^2+b^2}{1+16a^2b^2}
\end{equation}
Now substituting the values of these integrals in Eq.~(\ref{scbaA4}), and also retaining only the first harmonic of the oscillations we obtain
\begin{eqnarray}
K_{xx}\approx\frac{8e^2 ab}{\pi^2\hbar}\frac{a^2+b^2}{1+16a^2b^2}\bigg[1+2e^{-4\pi ab} \nonumber \\
\times \cos(2\pi (a^2-b^2) )\bigg]
\label{scbaA5}
\end{eqnarray}
Now substituting the expressions for $a$ and $b$ and keeping terms upto ${\cal O}(\hbar/\epsilon_F \tau)^2$ we obtain Eq.~(\ref{scba7}).

 \end{document}